\documentclass[a4paper]{article}

\newcommand{\eh}[1]{\,\mathrm{#1}}

\newcommand{\kyr}{\eh{kyr}}

\newcommand{\ergs}{\eh{erg\,s^{-1}}}

\newcommand{\mug}{\eh{\mu G}}

\newcommand{\ttt}[1]{\times10^{#1}}

\newcommand{\tin}[1]{_\mathrm{#1}}

\newcommand{\dg}{^{\circ}}

\renewcommand{\epsilon}{\varepsilon}

\newcommand{\hess}{H.E.S.S.}

\newcommand{\edot}{\dot{E}}

\newcommand{\fref}[1]{Fig.~\ref{#1}}

\usepackage{icrc2013}
\usepackage[english]{babel}

\title{A Population of Teraelectronvolt Pulsar Wind Nebulae in the H.E.S.S.
Galactic Plane Survey}

\shorttitle{Pulsar Wind Nebula Population in H.E.S.S.}

\authors{
S.~Klepser$^{1}$,
S.~Carrigan$^{2}$,
E.~de~O\~{n}a~Wilhelmi$^{2,5}$,
C.~Deil$^{2}$,
A.~F\"orster$^{2}$,
V.~Marandon$^{2}$,
M.~Mayer$^{1,3}$,
K.~Stycz$^{1}$,
and K.~Valerius$^{4}$
for the H.E.S.S. Collaboration.
}

\afiliations{
$^1$ DESY, Platanenallee 6, D-15738 Zeuthen, Germany \\
$^2$ Max-Planck-Institut f\"ur Kernphysik, P.O. Box 103980, D-69029 Heidelberg, Germany \\
$^3$ Institut f\"ur Physik und Astronomie, Universit\"at Potsdam, Karl-Liebknecht-Str. 24/25, D-14476 Potsdam-Golm, Germany \\
$^4$ Erlangen Centre for Astroparticle Physics (ECAP), Universit\"at
Erlangen-N\"urnberg, Erwin-Rommel-Str. 1, D-91058 Erlangen, Germany \\
\scriptsize{
$^{5}$ now at: Institut de Ci\`{e}ncies de l'Espai (IEEC-CSIC), 08193 Bellaterra, Spain
}
}

\email{stefan.klepser@desy.de}

\abstract{The most numerous source class that emerged from the H.E.S.S.
Galactic Plane Survey are Pulsar Wind Nebulae (PWNe). The 2013
reanalysis of this survey, undertaken after almost 10 years of observations,
provides us with the most sensitive and most complete census of gamma-ray PWNe
to date. In addition to a uniform analysis of spectral and morphological
parameters, for the first time also flux upper limits for energetic young
pulsars were extracted from the data. We present a discussion of the
correlation between energetic pulsars and TeV objects, and their respective
properties. We will put the results in context with the current theoretical
understanding of PWNe and evaluate the plausibility of previously
non-established PWN candidates.
}


\begin{document}
\maketitle

\section{Introduction}

The H.E.S.S. Galactic Plane Survey (HGPS) is the first survey of TeV sources
in the inner part of the Milky Way. After ten years of data taking and an
extensive reanalysis of all data, a full release of a source catalog and
skymaps is in preparation \cite{icrc_catalog}. This allows us to make a census of the most frequent
source class in this catalog, Pulsar Wind Nebulae (PWNe).

When massive stars explode in a core-collapse supernova, a neutron star can remain as
a leftover which
is often detected as a pulsar through its pulsations emitted in radio or other wavelengths. This pulsar has an outflow of
relativistic electrons and positrons which runs into a termination shock and forms a magnetised plasma around
the pulsar. This plasma bubble, which is the PWN, grows and shines for many
thousands of years and can reach
sizes of tens of parsecs across. The mechanisms by which it emits non-thermal
radiation are synchrotron (radio to sub-GeV gamma rays) and inverse Compton
radiation (GeV to TeV gamma rays).

The interesting questions that are addressed in the efforts to
measure and understand the nature of PWNe are how they evolve spatially and
spectrally with time, and how the injection of particles from the pulsar works
and evolves.
The evolution of a particular PWN strongly depends on the surrounding supernova
remnant (SNR), and both the SNR and PWN evolution depend on the surrounding
medium and other factors. In this study, where we compare all PWNe detected in the HGPS, we
can see the large scatter of PWN properties caused by these different local
conditions, while at the same time we investigate common trends in
evolution that can be established nevertheless.

\section{Correlation of TeV Sources and Young Pulsars}

From the many firmly associated PWNe in the online catalog
TeVCat\footnote{http://tevcat.uchicago.edu/}, and also from previous works on
the whole PWN population \cite{svenja_merida07, thesis_marandon,
thesis_mayer}, it can be considered an established fact that young energetic pulsars
tend to have a TeV PWN detectable with current instruments, while old, less energetic ones do not. This is also
what can be expected from the theoretical viewpoint because old pulsars inject
less and less particles and at some point are not able to compensate diffusion
and radiative losses anymore. These PWNe are still there, but too faint and/or
too extended to be detected with present instruments. 

\begin{figure}[t]
  \centering
  \includegraphics[width=0.5\textwidth]{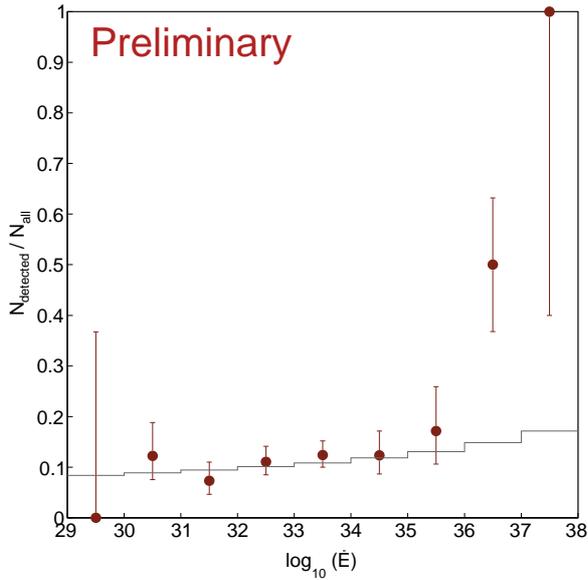}
  \caption{
Detection fraction of PMPS pulsars evaluated in bins of $\edot$. The black
histogram at the bottom is the expectation from randomised pulsar
populations, see text. The error bar of the highest-$\edot$ bin is large
because it is based on only two pulsars.
  \label{fig:detection_fraction}}
 \end{figure}

Using the new HGPS reanalysis, we update the study pursued in
\cite{svenja_merida07}. Figure \ref{fig:detection_fraction} shows the fraction of
pulsars with a TeV PWN detection as a function of the spin-down power $\edot$ of the
pulsar. The TeV detection is evaluated on the \hess\ survey skymap, and the
pulsar sample used is that of the Parkes Multibeam Pulsar Survey (PMPS)
\cite{pmps}. To estimate the fraction expected from chance coincidences, the
PMPS sample is randomised many times in galactic longitude and latitude
independently and for each bin and $\edot$. The expectation of chance
coincidences is displayed as a black line in \fref{fig:detection_fraction}.
The error bars on the data points are calculated with binomial statistics, as
described in \cite{thesis_svenja}. As in \cite{svenja_merida07}, we also find in this work that
pulsars with high $\edot$ are detected more often than can be expected by
chance, confirming the previous claims that these pulsars are more likely to
have a TeV PWN counterpart than the less energetic ones.

\begin{figure}[t]
  \centering
  \includegraphics[width=0.5\textwidth]{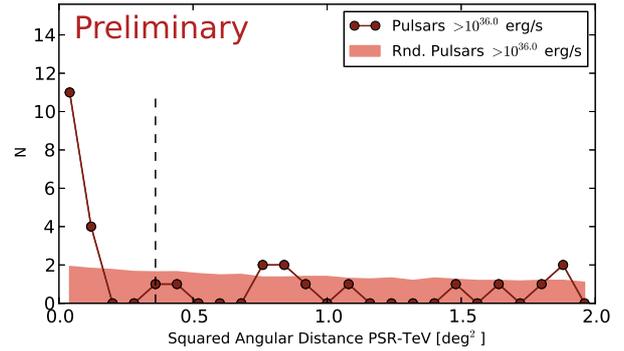}
  \caption{
Spatial correlation of high-$\edot$ PMPS pulsars and eligible HGPS source
candidates.
A clear correlation is seen as a peak at small angular distances, where very few chance coincidences
are expected. The dashed line indicates the preliminary
criterion to look for preselection candidates, see text.
}
  \label{fig:theta2}
 \end{figure}

In a similar fashion, this correlation can be demonstrated in a spatial
correlation plot (\fref{fig:theta2}), which shows the distribution of angular
distances $\theta$ between all PMPS pulsars and all TeV catalog source
positions. To evaluate the expected rate of chance coincidences, the same
distribution is again calculated with randomised pulsar populations, shown as a band. At correlation angles of
less than about $0.5\dg$, a strong peak is found.

\subsection{Candidate Preselection}

In the study of the PWN population, which will also evaluate the various
unconfirmed PWNe, a two-step candidate selection procedure is
applied. First, a very rough \textit{pre}selection is made, using rather
loose criteria in order not to miss a potential association. TeV sources that
are clearly identified as not being a PWN are excluded from this preselection. Secondly, these
preselected candidates are discussed in the context of the confirmed PWNe and model
predictions. They will be shown in a distinct color in the physics correlation
plots in Sec.~\ref{seq:conclusions}.

\begin{figure}[t]
  \centering
  \includegraphics[width=0.4\textwidth]{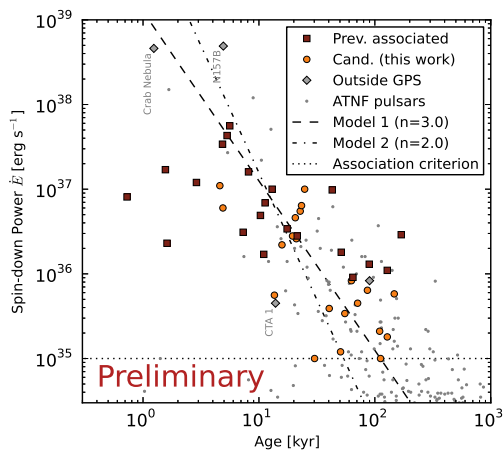}
  \caption{Spin-down power $\edot$ plotted against characteristic age of previously
identified PWNe (dark red squares) and preliminary preselection candidates in
this work (yellow circles). The grey diamonds are additional PWNe that are not in
the HGPS, and the grey dots are other ATNF pulsars for which no TeV
counterpart has
been detected.
}
  \label{fig:edot_age}
 \end{figure}

Figure \ref{fig:edot_age} shows a preliminary plot of the spin-down powers and
characteristic ages of the PWNe and candidates that fulfill the preselection
criteria in $\edot$ and angular separation $\theta$. As can be expected, the
confirmed PWNe are associated to young and energetic pulsars, whereas the
candidates are older and less energetic, making an identification with a TeV
source harder to argue.

\section{Conclusions on PWN Evolution}\label{seq:conclusions}

PWN evolution is mostly described to happen in three distinct stages: An initial phase
of free expansion within the ejecta of the supernova remnant (SNR) shell, a phase of interaction
and reverbations of PWN and SNR reverse shock, and finally a relic phase,
where pulsar, PWN and SNR might be disconnected from each other and a
secondary PWN bubble can form. Most of the known TeV PWNe are in either of the
first two phases.

\subsection{General Trends in PWN Modeling}

While the free expansion phase was often explored in theory and is comparably
well-understood, also thanks to the prominent Crab Nebula, the second phase,
where the PWN enters a complex interaction with the SNR reverse shock is still
relatively difficult to describe in a quantitatively realistic way. Most of
the works either strongly simplify the geometry of the system and/or are
done in numerical fashion, making them a case study rather than providing general
formulae. We therefore do not attempt to model a full synthetic population of
PWNe in the galaxy, but compare the phenomenological trends in the data we have with the common wisdom that can be
taken from present theory.

To understand the TeV spectral luminosity of a PWN it is essential to
model the electron population in bins of time \cite{martin, michael_model}.
Where required, we therefore use a simple, but time-dependent model similar to that outlined
in \cite{michael_model} and the formulae therein. To roughly characterise the
expected evolution of an average PWN, we assume an initial magnetic field of $50\mug$,
and two combinations of initial spin-down time scale $\tau_0$, initial $\edot$ and braking index $n$,
namely $\tau_0=0.5\kyr$; $5\ttt{39}\ergs$; $n=3$ (Model~1), and
$\tau_0=5\kyr$; $1\ttt{39}\ergs$; $n=2$
(Model~2). These modelings were chosen to represent two pulsar scenarios with different
decay characteristics, and $\tau_0$ and $\edot_0$ were adjusted for each given $n$ such that the
expected evolution curve in characteristic age appears correctly scaled on \fref{fig:edot_age}.
In addition to what is described in \cite{michael_model}, the model now
contains a free expansion phase in its radius evolution, which increases the
adiabatic losses in the early evolution. To compensate the efficiency loss
implied by that, we increased the lepton conversion efficiency from $0.3$ to
$1$.

\subsection{Extension Evolution}

The extension of a PWN is theoretically expected to develop as
$R\tin{PWN}\sim t^{1.2...0.3}$, with an index depending on the evolutionary state
\cite{gs,rc}. This radius $R\tin{PWN}$ refers to a sphere which the PWN bubble
is usually assumed to fill isotropically with electron/positron plasma. The
absolute scale of the radius evolution depends on the surrounding medium and
the SNR evolution and can therefore vary considerably.

The measurements available are radii defined as the width $\sigma$ of a
two-dimensional Gaussian function instead of a sphere radius. This $\sigma$ can only be measured if the
PWN is bigger than the minimum resolvable extension, which depends on the
exposure time and may generally be less than the gamma-ray point spread function.
Also, the source has to be sufficiently smaller than the radius of
the field of view, such that a background subtraction remains possible. For
\hess, this limits the range of detectable extensions to about $0.03\dg$ to
$0.6\dg$.

\begin{figure}[t]
  \centering
  \includegraphics[width=0.4\textwidth]{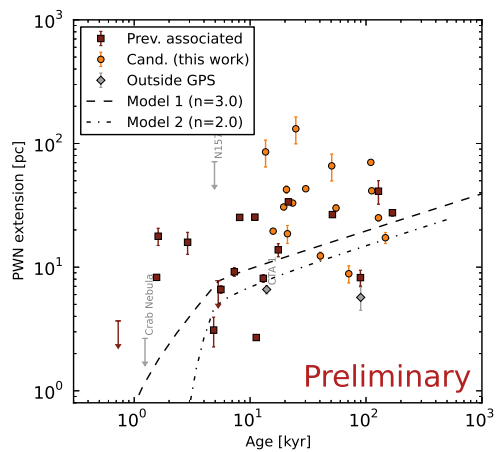}
  \caption{Dependence of measured PWN extensions on characteristic age. The
breaks in the models happen at the time when the PWN encounters the SNR reverse shock, which
is assumed to happen at $5\kyr$ (true age, as opposed to characteristic age which
is shown in the plot).}
  \label{fig:ext_age}
 \end{figure}

Despite this observational bias and theoretical uncertainty, a rough trend can
be observed in \fref{fig:ext_age}. The evolution of radii roughly matches the
expected $R\tin{PWN}\sim t^{0.3}$ for evolved PWNe that are in interaction with the SNR reverse shock. The model
curves in our figures take into account the conversion between true and characteristic age, see \cite{gs,rc}.

\subsection{Efficiency Evolution}

The TeV efficiency, defined as the ratio of TeV luminosity and pulsar
spin-down power, is often used to argue the plausibility of a PWN association.
It is, however, a difficult matter, because the TeV emission is produced by a
population of electrons that is summed up over the whole evolution of a PWN,
while the spin-down power is a momentary property of a pulsar. It is therefore
expected that this apparent efficiency of a PWN can increase with time or even
exceed unity.

\begin{figure}[t]
  \centering
  \includegraphics[width=0.4\textwidth]{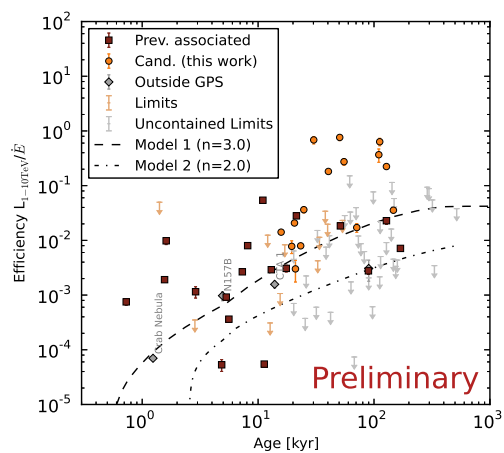}
  \caption{PWN efficiency vs. characteristic age. The yellow limits are likely
to enclose the respective PWN, whereas the grey limits are likely to limit
only the inner core of it, so the PWN is not fully contained.}
  \label{fig:eff_age}
 \end{figure}

Figure \fref{fig:eff_age} shows the evolution of efficiency with
characteristic age. Thanks to the many new upper limits we extract from the
HGPS skymap, it is clear that while some candidates may 
exceed an efficiency of $1$, most of the PWNe do not reach such a high
efficiency. From the modelings we show, one can see that the braking index $n$, which defines how
quickly $\edot$ decreases with time, probably plays an important role in
the efficiency evolution. A lower braking index of $n=2$ leads
to less outflow than $n=3$ and therefore less efficiency. On the other hand,
we also find that an even lower $n$ can lead to an artificial boost in efficiency by letting the
momentary $\edot$ drop very rapidly (not displayed here). A high efficiency may therfore either
indicate a productive particle generation, or just a fast decrease of the
pulsar $\edot$.

\section{Summary and Outlook}

We presented first excerpts of a study of the TeV PWNe population revealed by
the \hess\ Galactic Plane Survey. The study is to be completed using the final
catalog, and will contain a discussion of the PWN candidates and more
conclusions on the present theoretical understanding of pulsar wind nebulae.

\vspace*{0.5cm}
\footnotesize{{\bf Acknowledgment:}{The support of the Namibian authorities and of the University of Namibia
in facilitating the construction and operation of H.E.S.S. is gratefully
acknowledged, as is the support by the German Ministry for Education and
Research (BMBF), the Max Planck Society, the German Research Foundation (DFG), 
the French Ministry for Research,
the CNRS-IN2P3 and the Astroparticle Interdisciplinary Programme of the
CNRS, the U.K. Science and Technology Facilities Council (STFC),
the IPNP of the Charles University, the Czech Science Foundation, the Polish 
Ministry of Science and  Higher Education, the South African Department of
Science and Technology and National Research Foundation, and by the
University of Namibia. We appreciate the excellent work of the technical
support staff in Berlin, Durham, Hamburg, Heidelberg, Palaiseau, Paris,
Saclay, and in Namibia in the construction and operation of the
equipment.

}}

\end{document}